# Splitting of self-collimated beams in two-dimensional sonic crystals

Bo Li, Jun-Jun Guan, Ke Deng[1], and Heping Zhao

Department of Physics, Jishou University, Jishou 416000, Hunan, China

An easy-to-implement scheme to split self-collimated acoustic beams in sonic crystals (SCs) is proposed by introducing line-defects into SCs, by which an incoming self-collimated beam can be split into a 90˚-bended one and a transmitted one with an arbitrary power ratio by adjusting the value of defect size. An all-angle and wide-band splitting instrument is demonstrated with nearly perfect efficiency (more than 90%) for Gaussian beams at a wide range of incident angles. Splitting effect for a point source as input is also realized, in which two subwavelength images of the source, i.e., a bended one and a transmitted one, are formed by our designed splitting structure. Finally, the proposed one-to-two splitting scheme is generalized to one-to-N (N>2) cases by inserting more rows of line defects into the SC.

The richness of dispersion bands in sonic crystals (SCs) allows the manipulation of acoustic waves at the scale of wavelength in various unprecedented ways. This gives rise to many promising effects such as beaming [1-3], focusing [4-7] and imaging [8-11] of sonic waves, which act as elementary functionalities in on-chip solutions of acoustic application. To really implement these solutions, however, one still needs efficient routing of acoustic signals between different devices. Actually, in the integration of SCs onto a multifunctional chip, the self-collimation effect [12-16], by which acoustic signals can propagate in SCs with almost no diffraction along a definite direction, manifests itself as a potential candidate to perform this role. As a routing mechanism, self-collimation allows low-loss transmissions for a wide range of incident angles. It does not depend on physical channels to confine signals, thus gaining many advantages over conventional line-defect based routing mechanisms in which extra coupling loss and reflection loss are inevitable with the introducing of physical boundaries into SCs [17]. Moreover, profited from long flat equifrequency contours (EFCs), evanescent modes from an input source can also be transmitted by this routing mechanism; thus, self-collimation also possesses promising prospects in subwavelength applications [10]. However, implementation of this candidate suffers from the intrinsic inability of self-collimation to efficiently bend and split signals. Therefore, simple and efficient ways to bend and split self-collimated beams is urgently required for the SC integrated circuits.

---

[1]Corresponding author, e-mail address: dengke@jsu.edu.cn.

In our recent work [18], a high-efficiency sharp bending scheme for the self-collimation of acoustic waves has been proposed by simply truncating the SC. This scheme is based on the total internal reflection (TIR) of self-collimated signals at the interface between SC and matrix. In this paper, a splitting scheme for self-collimated acoustic signals is proposed by introducing a line-defect into the SC. As the combined results of TIR effect and tunneling effect at the line-defect, an incoming self-collimated beam can be split into bended and transmitted beams with an arbitrary power ratio by adjusting the value of defect size. This splitting technique is applied to realize a split imaging for a point source with resolutions beyond diffraction limit. Finally, the proposed $1 \times 2$ splitter is naturally generalized to $1 \times N (N > 2)$ cases by inserting more rows of line defect into the SC.

The two-dimensional SC employed here is composed of rubber-coated tungsten rods in water matrix, arranged in a square lattice. The outer and inner radii of the coating layer are $0.429a$ and $0.398a$ respectively, where $a$ is the lattice constant. The material parameters are $\rho = 19.3 \times 10^3 \text{kg/m}^3$, $C^L = 5.09 \times 10^3 \text{m/s}$ and $C^T = 2.8 \times 10^3 \text{m/s}$ for tungsten; $\rho = 1.3 \times 10^3 \text{kg/m}^3$, $C^L = 0.2 \times 10^3 \text{m/s}$ and $C^T = 0.04 \times 10^3 \text{m/s}$ for rubber;

$\rho = 1.0 \times 10^3 \, \text{kg/m}^3$ and $C^L = 1.49 \times 10^3 \, \text{m/s}$ for water. Here $\rho$ represents mass density, $C^L$ and $C^T$ represent the velocities of longitudinal and transverse waves, respectively. All results in this paper are calculated by the multiple-scattering theory (MST) [19,20].

Figure 1(a) gives the first band of the SC. Figure 1(b) presents several EFCs around the frequency 0.115. One can see that the EFCs around 0.115 are very flat along the ΓM direction and are much long compared to the EFCs of water [the dashed circle in Fig. 1(b) denotes the EFC of water at 0.115]. Therefore, if we put a source around the frequency 0.115 in water near one of the SC's ΓM interfaces, all of its propagating spatial harmonics and a wide range of evanescent ones will refract into the eigenmodes with wave vectors closed to the one denoted as $q$ in the Fig. 1(b), and then self-collimate along the ΓM direction in the SC structure [12-16]. These self-collimated beams will undergo a TIR at a crystal-water interface along the ΓX direction, as their wave vector components parallel to this direction lie outside the EFCs of water [as indicated by the dashed arrows in Fig. 1(b)]. Based on this TIR, we have realized a 90° bending technique for self-collimated beams by truncating the SC to introduce a ΓX-derationed interface [18]. As demonstrated in [18], nearly perfect routing of acoustic signals can be achieved in this sharp bending scheme, both for all the propagating and a wide range of evanescent spatial harmonics of a source. As an example, Fig 2(a) illustrates the bending effect for a vertically illuminated Gaussian beam at frequency 0.115.

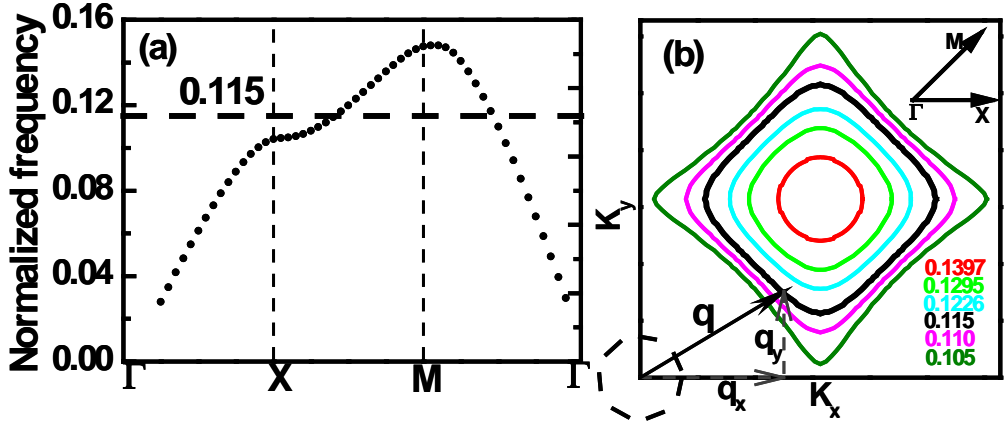

FIG. 1 (Color online) (a) The lowest band for SC considered here. (b) Several EFCs around 0.115. The dashed circle is the EFC of water at frequency 0.115.

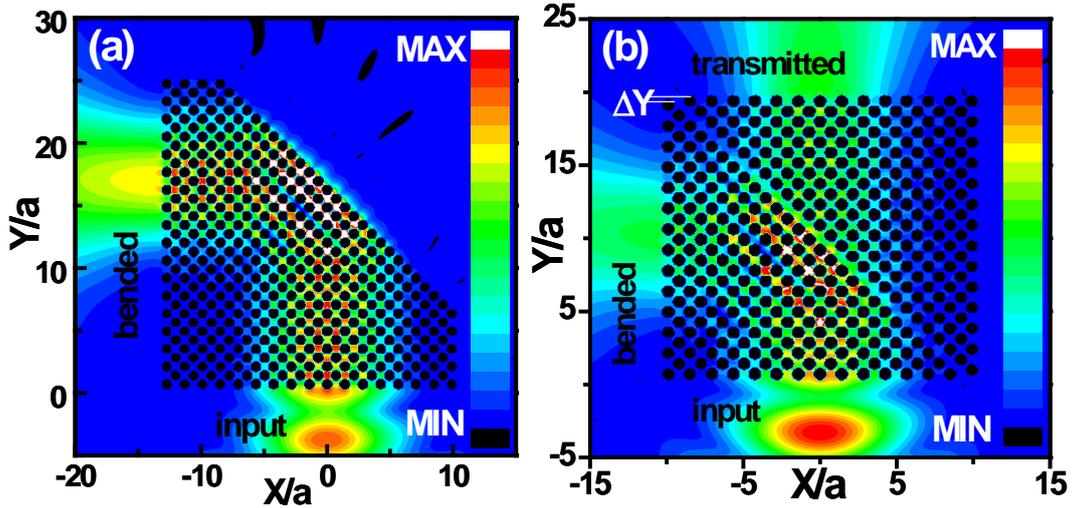

FIG. 2 (Color online) (a) Bending effect of self-collimated beams for a vertically illuminated Gaussian beam at frequency 0.115. (b) Splitting effect for a vertically illuminated Gaussian beam at frequency 0.115. Here the line defect is introduced by removing the right-upper portion of the self-collimating SC structure a little distance ΔY along the Y direction.

As one can see from Fig. 2(a), field amplitude decays very rapidly into water at the interface when the TIR of self-collimated beams occurs, becoming negligible at a distance within one lattice constant. However, if this interface is replaced by a ΓX-derationed water layer of finite thickness (i.e., a line-defect), a partially reflection will be expected due to the coupling of the two SC parts around the defect [21-24]. Therefore an incoming self-collimated beam can be split into reflected and transmitted ones at the introduced defect. Based on this partially reflection, a beam splitter is designed by removing the right-upper portion of the self-collimating SC structure a little distance ΔY along the Y direction as shown in Fig. 2(b). Splitting effect for a vertically illuminated Gaussian beam at frequency 0.115 is illustrated in Fig. 2(b), in which the self-collimated beam is separated into two parts, one is along the original propagation direction (the transmitted one), and the other is perpendicular to the original propagation direction (the bended one).

The transmitted part in Fig. 2(b) is caused by the coupling of Bloch modes separated by the interfaces, thus its intensity depends sensitively on the defect size. In Fig. 3(a) we give the power ratio between the bended and transmitted beams at frequency 0.115 versus the size of the line defect ΔY. Here the bended and the transmitted powers were normalized with respect to the input one and ΔY is varied from -0.2$a$ to 1.0$a$. The result clearly shows that an incoming self-collimated beam can be split into bended and transmitted ones with arbitrary power ratios by adjusting ΔY. It should be point out that the sum of bended and transmitted powers in Fig. 3 is less than the input one by about 10%. This is caused by the scattering loss of the SC structure.

The bending and transmitting efficiency as a function of frequency are shown in Fig. 3(b) for ΔY=0.3$a$, where the frequency is ranged from 0.100 to 0.130. We find that this splitting effect is of high-efficiency during the self-collimating frequency range as the sum of the bended and transmitted efficiency is always about 90%.

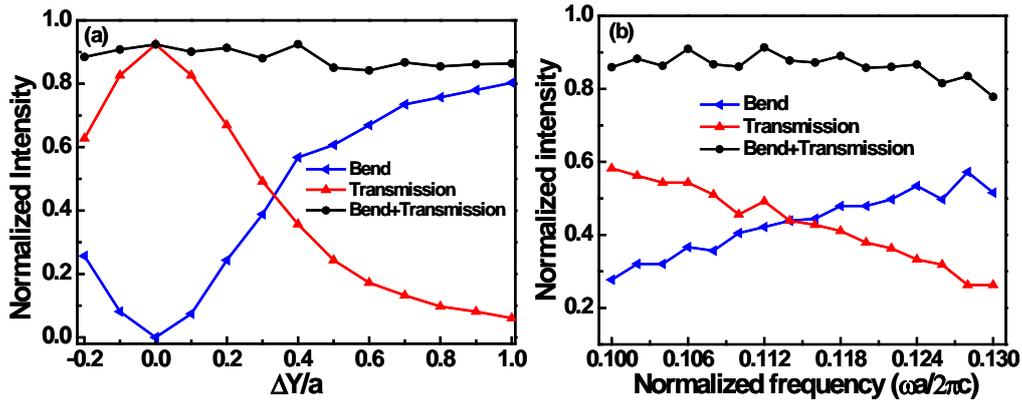

FIG. 3 (Color online) (a) Bended and transmitted powers which are normalized with respect to the input power as functions of ΔY. (b) Normalized Bended and transmitted powers as functions of frequency.

In Figure 4 we show the splitting effect for the case of oblique incidence. Pressure field distribution for a Gaussian beam at frequency 0.115 with incident angle $\theta = 35^o$ is shown in Fig. 4(a), which also exhibits a remarkable bending-transmitting effect. Figure 4(b) gives the splitting efficiency for Gaussian beams at frequency 0.115 with different incident angles. One can see that highly efficient splitting can be achieved for a wide range of incident angles. When $\theta > 50^o$, the efficiency falls sharply due to the rapidly increased reflections at the input port.

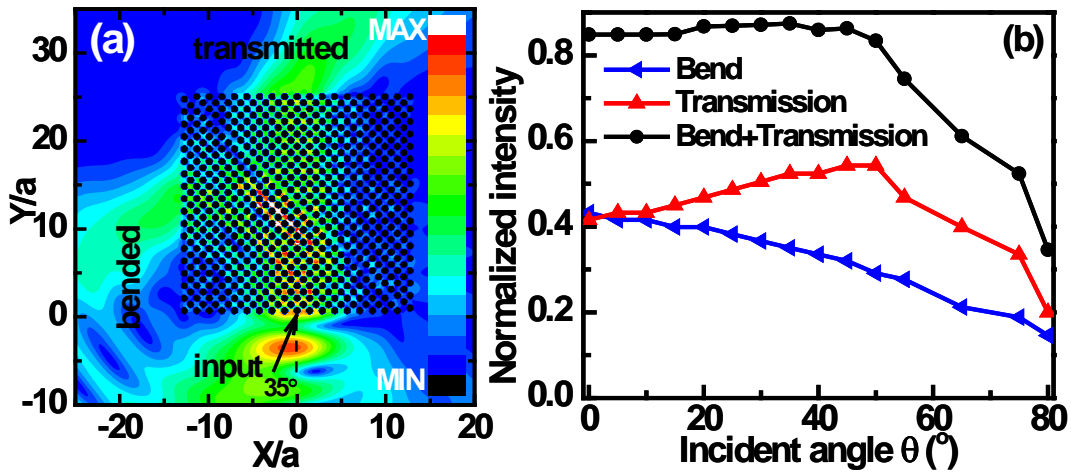

FIG. 4 (Color online) (a) Splitting effect for a Gaussian beam with incident angle $\theta = 35^o$. (b) Splitting efficiency for Gaussian beams with different incident angles.

Since the collimating effect holds for all propagating and a wide range of evanescent spatial harmonics of a source, splitting effects for a point source with subwavelength resolution can be expected. Figure 5(a) gives the pressure field distribution for a point source placed very close ($\sqrt{2}a/2$) to the under surface of the SC structure. One can see that two images of the source, i.e., a bended one and a transmitted one, are formed by our designed splitting structure. Figure 5(b) shows the full width at half maximum (FWHM) of the two images for different ΔY from -0.2$a$ to 0.7$a$. One sees that for the transmitting port, subwavelength imaging can be achieved when the line defect size is smaller than 0.5$a$, while for the bending port, image with subwavelength resolution can always be formed.

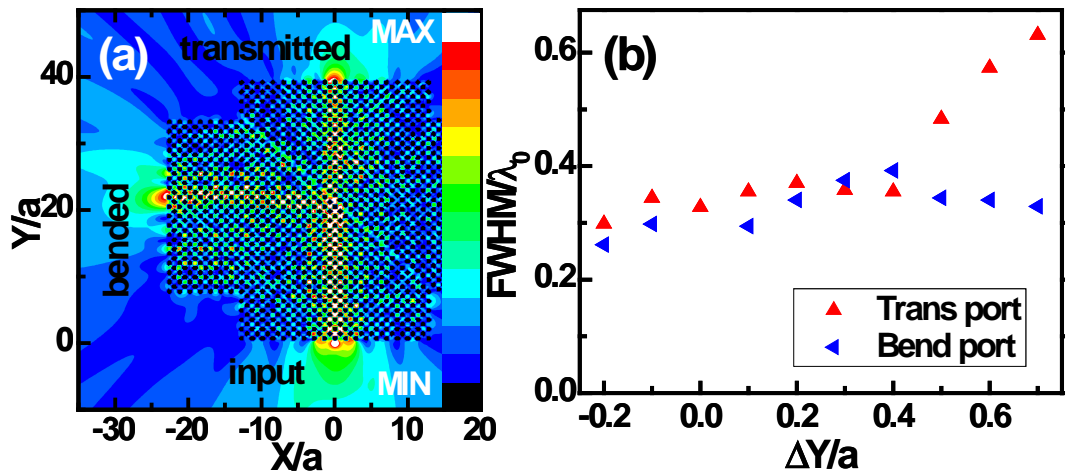

FIG. 5 (Color online) (a) Pressure field distribution of a point source placed $\sqrt{2}a/2$ away from the under surface of a splitting structure. (b) Full width at half maximum (FWHM) for images at transmitted and bended port for different ΔY.

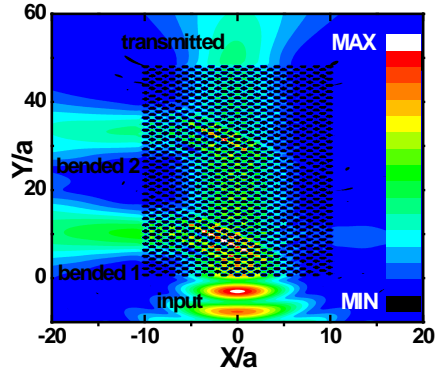

FIG. 6 (Color online) A one-to-three beam splitter composed of two line-defects.

This line-defect-based $1\times 2$ splitting mechanism can be easily generalized to $1\times N(N>2)$ cases by inserting more rows of line defect into the SC. Figure 6 demonstrates the splitting effect of a one-to-three splitter with a vertically illuminated Gaussian beam as input. Similar to the one-to-two case, the power of each output beam can also be easily controlled by varying the size of each defect.

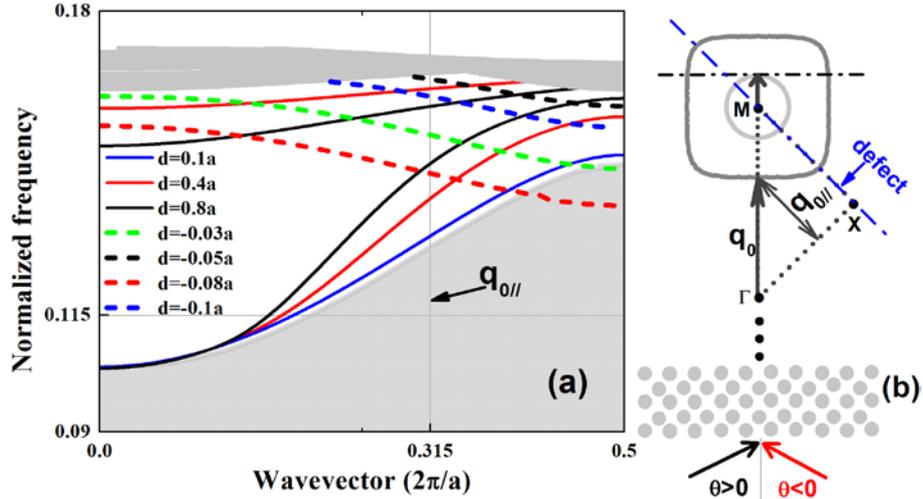

FIG. 7 (Color online) (a) Band diagrams for SCs with $\Gamma X$ derationed line-defects of different sizes $d$. Here $d=0$ refers to no defect. (b) The gray square-like curve around M point and the light gray cycle represents EFC of SC and water at frequency 0.115, respectively. The gray and blue dash-dot line represents the direction of water-SC incident interface and line-defect,

respectively. Here $q_0$ is the Bloch mode for incident angle $\theta = 0$.

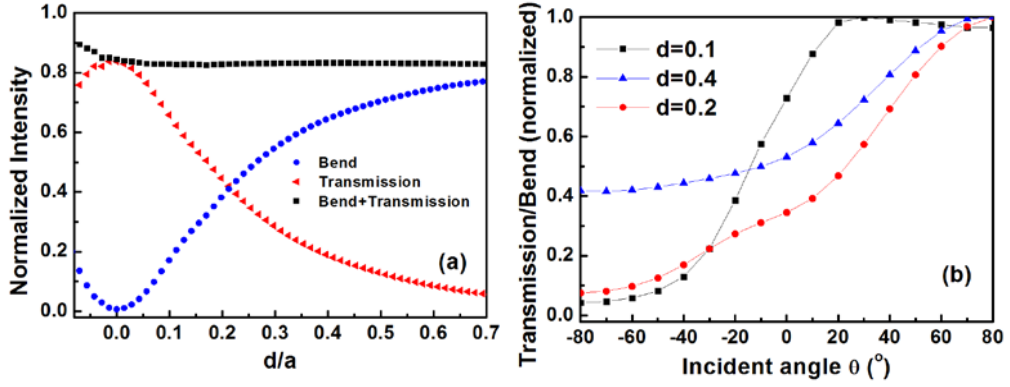

FIG. 8 (Color online) (a) Bended and transmitted powers normalized with respect to the input one as functions of $d$ for $\theta = 0$. (b) Transmission to bending ratios for different defect sizes as functions of incident angle.

Now we give some more detailed analysis to the mechanism of splitting. The Bloch modes in the two parts of SC are coupling by the defect modes of the introduced defect between the two parts, and the power ratio for the bended and transmitted beams is dependent on the coupling efficiency. Figure 7(a) gives the projected diagrams for SCs with a line-defect along the $\Gamma X$ direction, for different values of defect sizes $d$ (here $d = 0$ means no defect). The gray domains in the band diagram represent the continuum of the guide modes outside the band gap, and the curves represent the dispersion relations of defect modes. It is observed that for all defect sizes of $d < 0$, the frequency domains of defect modes are larger than 0.115. Therefore, these defect modes will not couple with the self-collimated one studied here, and we only need to consider the cases of $d > 0$. The Selection rule for coupling between the self-collimated Bloch modes $q$ and the defect mode can be deuced from the conservation of the projection of the k-vector onto the shared $\Gamma X$ direction modulo $2\pi / a$ [23,24]. In other words, in order for a self-collimated mode to couple to a defect one, the projection onto $\Gamma X$ of the reduced k-vector of the self-collimated mode [$q_{//}$ in Fig.7(b)] should be equal to the reduced k-vector of the defect mode. The coupling efficiency is closely relevant to the difference of k-vectors of the two modes. When the mismatch is large, the coupling efficiency becomes small. As can one see from the Fig.7(a), for the case of vertically illuminated Gaussian beam [$\theta = 0^o$, see Fig.2(b)], the projected k-vector of self-collimated mode onto $\Gamma X$ direction is $q_{0//} = 0.315$ at frequency 0.115. It is shown in the Fig.7(a) that, when the defect size increases, the k-vector of defect mode decreases and becomes more and more far away from $q_{0//}$. Thus, the couple efficiency is a decreasing function of the defect sizes. Figure 8(a) gives the normalized intensities for incident angle $\theta = 0^o$ as functions of defect size $d$. The transmission one decreases with the increase of $d$, which is in agreement with coupling theory discussed above. For a fixed $d$, when the incident angle increases, [ see the case of $\theta > 0$ in the Fig. 8(b)], one sees that the projected k-vector of self-collimated modes $d_{//}$ decreases, and become more close to the k-vector of defect mode, leading to the enhancement of coupling; when the incident angle decreases [the case of $\theta < 0$ in the Fig. 8(b)], $q_{//}$ increases and become more far away from the k-vector of defect mode, gives rise to the weakening of coupling. Figure 8(b) gives the transmission to bending ratio for different defect sizes as functions of incident angle. One can see that also the results are in agreement with the coupling theory.

In summary, we have proposed a splitting scheme for self-collimated acoustic signals by introducing line-defects into the SC. An incoming self-collimated beam can be split into bended and transmitted ones with an arbitrary power ratio by adjusting the value of defect size. We believe this easy-to-implement splitting mechanism will find applications in various multifunctional acoustic solutions integrated by different SCs devices.

**Acknowledgments**: This work is supported by the National Natural Science

Foundation of China (Grant No. 11104113 and No. 11264011), and Natural Science Foundation of Hunan province, China (Grant No. 11JJ6007), and Natural Science Foundation of Education Department of Hunan Province, China (Grant No. 11C1057).